\documentclass[useAMS,usenatbib]{mn2e}
\usepackage{natbib}
\usepackage{color,graphicx} 
\usepackage{amsmath}        
\usepackage{amsfonts}       
\usepackage{amssymb}        
\usepackage{wasysym}        
\usepackage{gensymb}        
\usepackage{framed}         
\usepackage{upgreek}
\usepackage{xspace}

\newcommand{\rxj}{RX~J0648.0$-$4418\xspace}

\author[E.F.~Keane]{E.F.~Keane$^{1,2}$ \\ $^{1}$ Centre for
  Astrophysics and Supercomputing, Swinburne University of Technology,
  Mail H30, PO Box 218, VIC 3122, Australia. \\ $^{2}$ ARC Centre of
  Excellence for All-sky Astrophysics (CAASTRO).} \date{\today}
\title[A radio search for \rxj]{A search for coherent radio emission
  from \rxj}

\begin{document}

\maketitle

\begin{abstract}
  \rxj is a compact star with a spin period of $13.2$~s. It is either
  the most rapidly rotating white dwarf known or a slowly rotating
  neutron star. Here we report on the first searches for coherent
  pulsar-like radio emission from \rxj, both for sporadic bursts and
  steady periodic emission. No such emission was detected, with our
  limits
%
%
  suggesting that no such mechanism is active. We further searched our
  data for fast radio bursts to a dispersion measure corresponding to
  a redshift of $\sim 12$. We did not detect any such events. 

\end{abstract}


\section{Introduction}
\rxj is a compact star in a $1.548$-day binary system with HD~49798,
an sdO5.5 subdwarf~\citep{tha70}. HD~49798 is bright at optical
wavelengths ($V = 8.287(8)$, \citealt{lu07}), where studies have
enabled measurement of the mass function of the
system~\citep{sl94}. \rxj is bright at X-ray wavelengths where the
spectrum is well fit by a blackbody ($kT_{\mathrm{bb}}^{\infty}
\approx 39$~eV, $T_{\mathrm{bb}}^{\infty} \approx 0.45$~MK and
$R_{\mathrm{bb}}^{\infty} \approx 20$~km) plus power-law (photon
index, $\Gamma\sim 2$) combination~\citep{mte+09}. Modelling the
observed X-ray eclipse reveals both the inclination angle of the
system and the radius of HD~49798, as the distance is known to be
$650\pm100$~pc~\citep{ks78}. Both of these, combined with the mass
function, provide the mass for each star:
$1.50(5)\;\mathrm{M}_{\astrosun}$ for HD~49798 and
$1.28(5)\;\mathrm{M}_{\astrosun}$ for \rxj. Furthermore, \rxj shows
X-ray pulsations every $13.2$~s~\citep{isa+97}, which is interpreted
as its spin period, and the spin period derivative is $< 6 \times
10^{-15}$~\citep{mlt+13}.

The mass, spin period, spin period derivative, temperature and
blackbody radius are all consistent with \rxj being either a white
dwarf (WD) or a neutron star (NS). If \rxj is a WD it is the most
rapidly spinning one known, $\sim 2.5$ times faster than the second
fastest rotator AE Aqr ($P=33$~s, \citealt{pat79}). If it is a NS it
is $\sim 1.5$ times slower than PSR~J2144$-$3933, the slowest NS known
to show pulsed radio emission ($P=8.5$~s, \citealt{ymj99}). The
inferred rotational energy loss rate is $\lesssim 10^{22}$~W
($\lesssim 10^{28}$~W) if it is a WD (NS). In both scenarios there is
the possibility that \rxj is emitting bursts of coherent radio
emission, either as a so-called WD pulsar~\citep{mrr12} or as a NS
with an `almost-dead' radio pulsar mechanism~\citep{km11}. Despite
this, there have been no searches performed to try to detect this
radiation. In this paper we describe such a search. In \S~2 we
describe our observations, before detailing the different search
methods employed in, \S~3, as well as a Fast Radio Burst (FRB) search
which is possible `for free'. We provide our conclusions and a
discussion in \S~4.

\section{Observations} 
On October 23rd 2013, \rxj was observed for $548$ minutes using the
64-metre Parkes Radio Telescope in New South Wales, Australia. The
telescope was centred on the position of the source:
$\alpha_{\mathrm{J2000}}=06\mathrm{h}48\mathrm{m}04.700\mathrm{s}$,
$\delta_{\mathrm{J2000}}=-44\degree18^{\prime}58.44^{\prime\prime}$~\citep{lee07}. The
20-cm multi-beam receiver~\citep{swb+96}, which received orthogonal
linear polarisations, was used in combination with the
Berkeley-Parkes-Swinburne Recorder backend~\citep{kjs+10}. A bandwidth
of $400$~MHz, about a central frequency of $1382$~MHz, was Nyquist
sampled, channelised to $1024$ frequency channels, and then integrated
by a factor of $25$ for a sampling time of $64\;\upmu\mathrm{s}$. The
polarisations were summed to produce total intensity (Stokes
\textit{I}), and the data samples were written to disk as 2-bit
numbers. To mitigate radio frequency interference (RFI) using
coincidence testing, and to perform a commensal search for FRBs, data
were recorded using all 13 beams of the multi-beam receiver.

\section{Data Analysis \& Results}
To mitigate RFI two data cleaning steps were performed. (i) The $158$
frequency channels where there is known RFI due to powerful
transmissions from geostationary communications
satellites~\citep{kjs+10} were set to zero in value. (ii) The time
samples wherein the total power was seen to be highly correlated
across all 13 beams (characteristic of terrestrial interference),
judged using the method described in \citet{kbb+12}, were removed.
\textit{A priori} we do not know whether it will be easier to identify
any putative radio emission in a single pulse search or a periodicity
search.
It is straightforward to show that the ratio of the S/N ratios for
these searches is $r =
(2/\sqrt{N})(S_{\mathrm{peak}}/S_{\mathrm{ave}})$, where $N$ is the
number of pulse periods during the observation.
It is clear that, in the long-term, the $\sqrt{N}$ term dominates so
that a periodicity search is more effective in the large-$N$
limit. For most pulse amplitude distributions seen in pulsars there is
a `sweet spot' at lower $N$-values, where the single pulse search can
be more effective by an order of magnitude~\citep{kea10a}. The pulse
amplitude distribution, number of pulse periods observed and
brightness of the pulsar (or equivalently the sensitivity of the
instrument) together dictate the relative effectiveness of both search
methods. We employed both searches, as we describe below.


\subsection{Single Pulse Search}\label{sec:sp_search}
The data were searched for isolated bursts of radio emission using
\textsc{heimdall}\footnote{\texttt{http://sourceforge.net/projects/heimdall-astro/}},
a GPU code developed for realtime searches of pulsar survey data
(Barsdell et al., in prep.). The data were dedispersed to 1018 trial
dispersion measures (DM) in the range
$0-100\;\mathrm{pc}\,\mathrm{cm}^{-3}$. The estimated DM for the
source, according to the NE2001 model of the Galaxy's electron density
content, is $9.85\;\mathrm{pc}\,\mathrm{cm}^{-3}$~\citep{cl02}. This
model, although generally reliable to the $\sim 20\%$ level, can be
wrong by as much as a factor of $2$ along specific lines of sight
(Deller et al. 2009, Bannister et al. 2014).\nocite{dtbr09,bm14}
The wide range of DM trials searched was
chosen as it was considered extremely unlikely that the NE2001
prediction could be incorrect by more than a factor of $10$. We note
that the maximum Galactic DM contribution predicted for this line of
sight is $77.3\;\mathrm{pc}\,\mathrm{cm}^{-3}$. The spacing of our DM
trials was chosen using the prescription of \citet{lev12} with a ``DM
tolerance'' of 1.05, so our DM coverage is at a much higher resolution
than is typical for pulsar searches of large
surveys~\citep{kjs+10}. 

Our number of statistical trials is large, as we recorded $15625$
samples per second for $9.1$ hours, searching each of these $1018$
times. We searched for a range of pulses widths from $1$ to $2^{12}$
times the raw sampling time, and this repeated searching further
doubled the number of statistical trials~\citep{cm03}. In total the
number of trials is just over $10^{12}$, implying that we can expect
to find $\sim1$ event with S/N of $7$ by chance in our searches.
Due to the dish and receiver geometry it is impossible for a boresight
astrophysical signal to appear in more than $3$ beams. Thus we
filtered out such multi-beam events, which are likely to be
terrestrial in nature.

We then examined standard single pulse search diagnostic plots for
significant events in the DM range
$1-100\;\mathrm{pc}\,\mathrm{cm}^{-3}$; events peaking below
$1\;\mathrm{pc}\,\mathrm{cm}^{-3}$ are likely terrestrial. For events
with $W\leq 2^{10}t_{\mathrm{samp}} = 65.536\;\mathrm{ms}$, only six
events with $S/N\geq 8$ (a more realistic threshold given imperfect
RFI excision) were evident, which were all seen to be clearly
terrestrial upon inspection of their frequency-time behaviour; a
genuine astrophysical signal shows a characteristic $t(\nu) \propto
\nu^{-2}$ `sweep'. For events with $2^{10} < W/t_{\mathrm{samp}} \leq
2^{12}$, the width of a broadband signal is comparable to or larger
than the dispersion delay across the entire band, $t_{\mathrm{DM}}
\approx
131.2\,(DM/100\;\mathrm{pc}\,\mathrm{cm}^{-3})\;\mathrm{ms}$. It is
equivalent to note that the ``DM width'' becomes much wider than the
entire DM range searched (see \citealt{cm03}, Equation 4). A number of
such pulses were detected, but none with the frequency-time signature
or broad range in detected DM values one would expect from an
astrophysical signal. Wider pulses are ever less distinguishable from
zero-DM (i.e. terrestrial) signals, especially at the detection
threshold. As expected, the number of detections for each boxcar width
is approximately constant up to $2^{10}$, then increases for the
widest boxcars when wide zero-DM signals become detected at
essentially all DM values. Also as expected, there are excess events
detected, at all DM values and in all beams, for the width
corresponding to the $50$-Hz mains signal. A conservative 10-$\sigma$
limit on burst events can be calculated using the well-known
radiometer equation~\citep{lk05}.
This yields $S_{\mathrm{SP,10\sigma}} =
120\,/\sqrt{W/10\,\mathrm{ms}}\;\mathrm{mJy}$, where $W$ is the pulse
width. This corresponds to a radio pseudo-luminosity of
$<0.051\;\mathrm{Jy}\,\mathrm{kpc}^2 = 4.8\times 10^{11}
\;\mathrm{W}\,\mathrm{Hz}^{-1}$. This is an order of magnitude fainter
than the lowest luminosities detected in coherent radio bursts from
the sporadically emitting long-period NSs~\citep{kle+10}, and thus a
very constraining limit suggesting that a similar burst mechanism is
not present in \rxj.

\subsection{Periodicity Search}
The rotation period of \rxj is known to high precision~\citep{mlt+13}.
Nonetheless we performed a search in period, in the range $13.0$ to
$13.4$~s, far in excess of the period uncertainty. This search used
the Fast Folding Algorithm, which is superior to Fourier Transform
searches for periodicities $\gtrsim 1$~s~\citep{kml+09}, and
considered a duty cycle range of $\sim 0.01\%$ to $\sim 8\%$. The same
DM trials as per \S~\ref{sec:sp_search} were used. Our number of
statistical trials is $\sim 1.6 \times 10^9$, less than for the single
pulse search. Thus there is an expectation of $\sim 1$ signal with S/N
$\gtrsim 6$ simply by chance.
However, due to both chance time alignments of imperfectly excised RFI
events at integer multiples of our trial periods and the red noise
seen in the data (the latter having no influence on the single pulse
search), $8$ is a more realistic threshold. No signals with S/N$>8$
with the expected signatures of a genuine astrophysical signal were
observed. Matching our conservative approach from
\S~\ref{sec:sp_search}, we place a flux density limit on periodic
emission of $S_{\mathrm{fold, 10\sigma}} = 29 \;\mathrm{\upmu Jy}\; ,$
for an indicative duty cycle, $\delta$, of $1\%$; the sensitivity
scales as $\sqrt{\delta/(1-\delta)}$. This flux density limit is at
the level of the deepest limits on unidentified \textit{Fermi}
objects~\citep{bgc+13}. The corresponding radio pseudo-luminosity
limit is $1.2\times 10^{-5}\;\mathrm{Jy}\,\mathrm{kpc}^2 = 1.2\times
10^8\; \mathrm{W}\,\mathrm{Hz}^{-1}$.

\subsection{Fast Radio Burst Search}
\rxj is quite far off the Galactic plane ($l = 253.7065\degree$,
$b=-19.1409\degree$) and the product of observing time and field of
view for our observations is reasonably large at 
$5.4\;\mathrm{hour} \, \mathrm{deg}^{2}$. The FRB rate is
approximately one per 200 hours in the intergalactic volume probed by
the High Time Resolution Universe (HTRU)
survey~\citep{tsb+13}\footnote{We note however that the error on this
  number is very large. With $95\%$ error bars the $4$ events detected
  in \citet{tsb+13} should be taken as
  $4^{+5.1}_{-2.6}$~\citep{geh86}}. The HTRU search probed events up
to a DM of $2000\;\mathrm{pc}\,\mathrm{cm}^{-3}$. We searched our data
to a DM of $10,000\;\mathrm{pc}\,\mathrm{cm}^{-3}$. Subtracting a
reasonable estimate for the combined Milky Way plus host galaxy DM
contributions of $500\;\mathrm{pc}\,\mathrm{cm}^{-3}$ from both
searches, and using the DM-z relation of \citep{ioka03} we deduce the
relative volumes probed. If FRBs were standard candles visible to
(say) $z\sim 3$
then we might expect $\sim 1$ event during our observations. If FRBs
have a wide luminosity distribution then our high-DM search probes
even more volume. We did not detect any FRBs in our search. This
implies either (i) we have been unlucky as we are dealing with small
number statistics and a highly uncertain rate estimate, (ii) FRBs are
standard candles but with luminosities such that they are not
detectable out to $z\sim3$, or (iii) FRBs have a wide luminosity
distribution such that some are detectable to high redshift, but have
a much lower rate than estimated. We speculate that a combination of
the first two possibilities seems most likely.

\section{Discussion \& Conclusion}
\textit{Is \rxj a NS?} It does not show any periodic or bursting radio
emission down to 
quite stringent limits. It seems that such mechanisms are not in
operation, or that the beaming geometry of such emission is
unfavourable, although, as there are no pulsar beaming fraction
measurements for $P\gtrsim 3$~s~\citep{tm98}, it is difficult to make
an estimate of this beaming fraction. The possibility remains that it
is a slow pulsar whose radio emission mechanism is just extremely
intermittent. Our observations limit the emission of any detectable
burst to no more often than once every $\sim 10^5$ rotation
periods. For its `characteristic age' of $P/(2\dot{P})\gtrsim 35$~Myr
its temperature is too hot to be explained by standard NS
cooling~\citep{kmk+13} but is easily explained by accretion. The mass,
at $1.28(5)\;\mathrm{M}_{\astrosun}$, is on the light side of the
observed NS mass distribution, when corrections for the system binding
energy and any accreted mass are considered. Thus, any formation
scenario may need to invoke an electron capture
supernova~\citep{plp+04} in identifying a viable formation
scenario. The inferred NS magnetic field strength of $\lesssim
10^{13}$~G is sufficiently high that X-ray absorption lines might be
expected, but none are reported.

\textit{Is \rxj a WD?} \citet{mte+09} argue that, based on the X-ray
spectrum, a WD is a better fit, but that a NS is not ruled out. Our
observations probe coherent, non-thermal radio emission only, so
cannot limit the actual temperature of the star but only the
brightness temperature; our pseudo-luminosity limit corresponds to a
brightness temperature limit of $\lesssim 10^{20}$~K ($\lesssim
10^{16}$~K) for single pulse (periodic) emission. However if the
fractional radio energy loss rate is the same in WDs as it is for NSs,
our WD luminosity limit is thus $\sim 10^6$ times more
constraining. The inferred WD magnetic field strength is $\lesssim
10^{7}$~G, which would have no impact on the X-ray spectrum. We note
that if radio-emitting magnetars are in fact WDs, and \rxj were
similar, we would expect it to be visible in the frequency range we
observed.

\section*{Acknowledgements}
{\small EFK thanks Mark Purver, Willem van Straten, Matthew Bailes and
  the anonymous referee, for helpful comments which have improved the
  quality of this paper. This work used the gSTAR national facility
  which is funded by Swinburne and the Australian Government’s
  Education Investment Fund. The Parkes Radio Telescope is part of the
  Australia Telescope National Facility, which is funded by the
  Commonwealth of Australia for operation as a National Facility
  managed by CSIRO.
EFK acknowledges the support of the Australian Research Council Centre
of Excellence for All-sky Astrophysics (CAASTRO), through project
number CE110001020.}

\bibliographystyle{mnras}


\end{document}